\documentclass[preprint]{aastex}

\newcommand{\teff}{$T_{eff}$}
\newcommand{\grav}{log($g$)}
\newcommand{\etall}{{\it et al.\/}}

\newcommand{\cband}{C$_2$}

\begin{document}

\title{The Frequency of C-Rich Extremely Metal
Poor Stars\altaffilmark{1}}

\author{Judith G. Cohen\altaffilmark{2}, 
Stephen Shectman\altaffilmark{3} 
Ian Thompson\altaffilmark{3},
Andrew McWilliam\altaffilmark{3},  
Norbert Christlieb\altaffilmark{4},
Jorge Melendez\altaffilmark{2},
Franz-Josef Zickgraf\altaffilmark{4},
Solange Ram\'{\i}rez\altaffilmark{5}
\& Amber Swenson\altaffilmark{2} }

\altaffiltext{1}{Based in part on observations obtained at the
W.M. Keck Observatory, which is operated jointly by the California 
Institute of Technology, the University of California, and the
National Aeronautics and Space Administration.}

\altaffiltext{2}{Palomar Observatory, Mail Stop 105-24,
California Institute of Technology, Pasadena, Ca., 91125, 
jlc@astro.caltech.edu}

\altaffiltext{3}{Carnegie Observatories, 813 Santa
Barbara Street, Pasadena, Ca. 91101}

\altaffiltext{4}{Hamburger Sternwarte, Universit\"at
Hamburg, Gojenbergsweg 112, D-21029 Hamburg, Germany}

\altaffiltext{5}{Spitzer Science Center, Mail Stop 100-22,
California Institute of Technology, Pasadena, Ca., 91125}

\begin{abstract}
We demonstrate that there are systematic scale errors in
the [Fe/H] values determined by the Hamburg/ESO Survey
(and by the HK Survey by inference) for certain extremely metal
poor (EMP) highly C-enhanced giants.  The consequences of these scale errors 
are that a) the fraction of carbon stars at extremely low metallicities 
has been substantially overestimated in several papers in the recent 
literature b) the number of EMP stars known is somewhat 
lower than has been quoted in the recent literature c) the yield for 
EMP stars by the HK and the HES Survey is somewhat lower
than is stated in the recent literature.  A preliminary
estimate for the frequency of Carbon stars among the giants in the HES
sample with $-4 <$ [Fe/H] $ < -2.0$ dex is 7.4$\pm2.9$\%,
and for C-rich giants with [C/Fe] $\ge$ +1.0 dex is 14.4$\pm4$\%.

We rely on the results of an extensive set of detailed
abundance analyses of stars expected to have
[Fe/H] $\le -3.0$ dex  selected from the Hamburg/ESO Survey
to establish these claims. These analyses of $\sim$50 HES candidate
extremely metal poor stars have been carried out in
as homogeneous a manner as possible.  Here we present the key results of 
detailed abundance analyses of 14 C-stars selected in this way 
About 80\% of such C-stars show highly enhanced Ba as well,
with C enhanced by a factor of about 100, and [Ba/C] roughly Solar.
These stars often show prominent lead lines, and presumably are the
remnants of the secondary in a mass transfer binary system where the 
former primary was an AGB star, which transferred substantial mass at that 
evolutionary stage.  The remaining 20\% of the C-stars do not show an 
enhancemement of the s-process neutron capture elements around the Ba peak. 
They tend to be the most metal-poor stars studied.  We suggest
that they too result from mass transfer across a binary system. 
(published abstract will be shorter due to space limitations)

\end{abstract}

\section{The Fraction of Carbon Stars Among Extremely Metal Poor Stars}

We are engaged in a large scale project to find additional
EMP stars in the halo of our galaxy by mining the database of the
HES. The normal procedures outlined by \cite{christlieb03}
to isolate EMP stars from the
candidate lists produced by the HES were followed.   The candidates
were vetted via moderate resolution spectroscopy
at large telescopes
to eliminate the numerous higher abundance interlopers.
Most of the follow up spectra for the stars discussed here were obtained
with the Double Spectrograph  on the Hale Telescope
at Palomar Mountain.  We intend to observe all candidates to the magnitude
limit of the HES (B $\sim$ 17.5) in our fields; with $\sim$1600 moderate
resolution spectra in hand from campaigns at Palomar and at the
Las Campanas Observatory, observations are now
complete in $\sim$990 deg$^2$, complete to B=16.5 in
an additional $\sim$700 deg$^2$,
and approaching completion in the remaining fields.

These follow up spectra are used to determine an accurate measure of
the metallicity of the star via a combination of strength of absorption in
H$\delta$ (determining \teff) and in the
Ca~II line at 3933~\AA\ (the KP index), which then determines
[Fe/H].  The calibration between the
strength of the indices
and resulting metallicity [Fe/H](HES) adopted by the HES is
described in \cite{beers99} and is essentially identical to that
used by the HK Survey until recently; the latest updates to the algorithm
as used by the HK Survey are described in \cite{rossi05}.

We have found that there are systematic scale errors in
the [Fe/H] values determined by this algorithm for certain EMP
highly C-enhanced giants.  As is shown in \cite{cohen_apjl},
these scale errors act to make certain C-stars
appear more metal poor by a factor of $\sim$10 than would be inferred from
a detailed abundance analysis of their high dispersion spectra.
These problems appear to arise due to molecular
absorption of CH and CN in the specific continuum bandpasses used to measure the
KP and the H$\delta$ line indices from which [Fe/H] is inferred.
This makes both of these lines appear weaker, and
hence the inferred Fe-abundance of such a C-star
is underestimated.  This effect is largest for the cooler C-stars (\teff $\sim$5100~K)
with high C-enhancement and high N/C ratios.  Normal-C giants
and the warmer C-stars are not affected; [Fe/H](HES) and [Fe/H](HIRES] are
in good agreement for them.

We rely on the results of an extensive set of detailed
abundance analyses of stars expected to have
[Fe/H] $\le -2.9$ dex  selected from the Hamburg/ESO Survey
to establish these claims.
We have obtained and analyzed spectra with HIRES \cite{vogt94}
at the Keck I Telescope of  $\sim$60 HES candidate
EMP stars.
\cite{cohen_cabund} will present abundance analyses for 14 of the 16
known Carbon stars from this HES sample.   Fifteen
C-normal giants have been analyzed to date as well, while
\cite{cohen04} studied a large sample of candidate
EMP dwarfs from the HES.

The first consequence of these scale errors is that
the fraction of carbon stars at extremely
low metallicities has been substantially overestimated
in several papers in the recent literature.
Our operational definition of a C-star is one
whose spectrum shows bands of \cband.
If no \cband\ bands are detected, but [C/Fe] $ > 1$ dex, we denote a
star to be C-enhanced.  The strength of the \cband\ bands will be
a function of \teff,  $\epsilon$(C), and to a lesser extent,
\grav\ and [Fe/H].
Also we denote stars with \teff\ $ > 6000$~K as ``dwarfs'', while all cooler
stars are called ``giants''.

\begin{figure}
\includegraphics[width=3.4in]{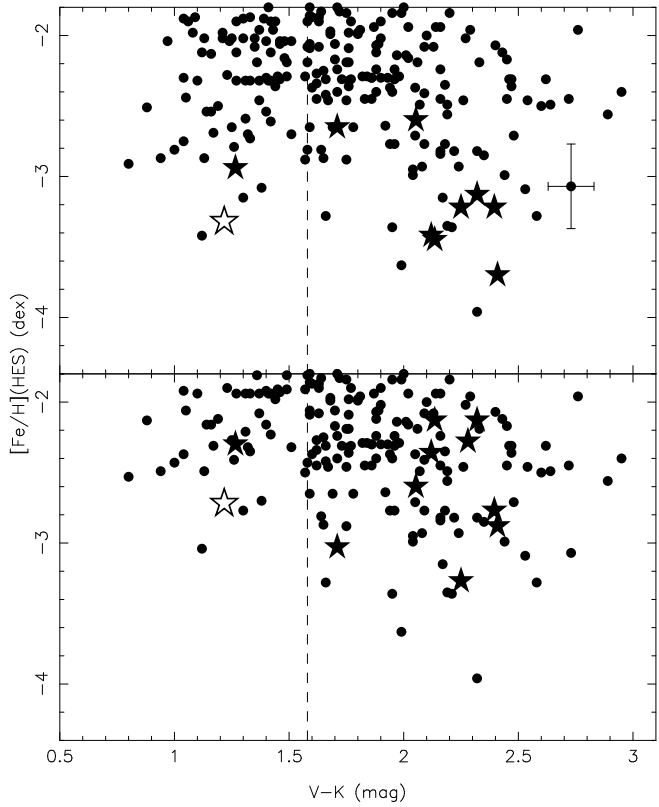}
  \caption{A plot of [Fe/H](HES) versus V-K for a sample of 489 EMP
candidates from the HES with moderate resolution
spectra from the Double Spectrograph at the Hale Telescope
(filled circles, limited to stars with [Fe/H](HES) $< -1.8$ dex).
The C-stars
are from this sample are indicated by filled stars; the C-enhanced star is
shown as an open star.  In the upper panel
the C-stars are plotted at their [Fe/H](HES) values, while in the lower
panel, at their [Fe/H](HIRES) values.
The C-normal dwarfs have also been shifted towards higher
[Fe/H] by 0.38 dex in the lower panel.
A typical error for a EMP giant with normal
C is shown for a single star in the upper panel.
The vertical dashed line separates the giants
from the dwarfs.}\label{fig_feh_vmk}
\end{figure}

Even though the cooler C-stars comprise only a small fraction of the
most metal poor stars, an underestimate of a factor of
$\sim$1 dex in their [Fe/H](HES) will have significant effects.
Fig.~\ref{fig_feh_vmk} shows [Fe/H](HES) versus V-K for a sample of 489 EMP
candidates from the HES with moderate resolution
spectra from the Double Spectrograph at the Hale Telescope.
The known C-stars and the C-enhanced star with HIRES
analyses from this sample are indicated.
In the upper panel, the C-rich stars are plotted
at their [Fe/H](HES) values, while in the lower panel
they are plotted at their [Fe/H](HIRES) as determined from detailed
abundance analyses.   Although at their
nominal Fe-metallicities the C-stars dominate the population of the
giants below [Fe/H](HES) $-3$ dex,
using the results from analysis of high resolution spectra in the lower
panel the frequency of C-stars
among the most metal poor EMP stars is  reduced by a factor of 2.5.  The
C-star frequency for the Palomar sample of
HES giants with [Fe/H](HES) $< -2.0$ dex is
7.4$\pm2.9$\%.
Adding in the fraction of C-enhanced stars
among giants with [Fe/H](HES) $< -2.0$ dex we establish below
of 6.5$\pm2.7$\%,
one obtains a total fraction of C-rich stars with
[C/Fe] $> +1.0$ dex of 14$\pm4$\% among our HES EMP sample with
[Fe/H](HES) $< -2.0$ dex.

The metallicity distribution function is very sharply declining
among halo stars at the lowest Fe-metallicities.
Thus the systematic errors we have found in the calibration of the HES and
by inference the HK metallicity scale of \cite{beers99}, at least
until quite recently (see Rossi \etall\ 2005)
will also lead directly to systematic
overestimates of the number of EMP stars and of the yield
for EMP stars by these two major surveys.  More details of this
work can be found in \cite{cohen_apjl}.

\section{Extremely Metal Poor C-Enhanced Stars}

\begin{figure}
\includegraphics[width=3.4in]{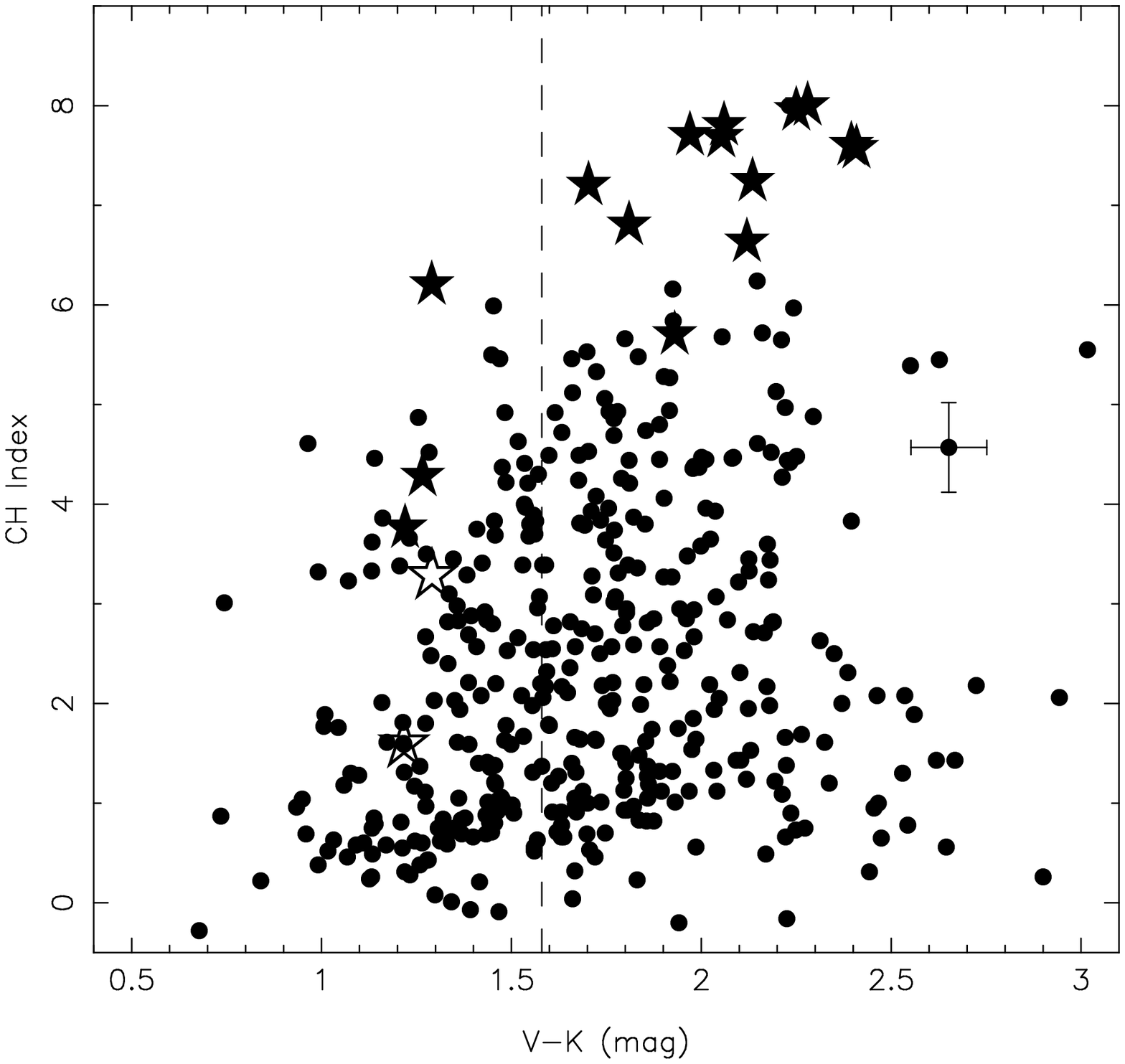}
  \caption{The CH index (GP) is shown as a function of V-K for the
full P200 sample of EMP candidates from the HES.
The known C-stars are shown as filled stars, the known C-enhanced stars as
open stars.
A typical error for a EMP giant with normal
C is shown for a single star.
The vertical dashed line separates the giants
from the dwarfs.}\label{fig_ch_vmk}
\end{figure}

\begin{figure}
\includegraphics[width=3.6in]{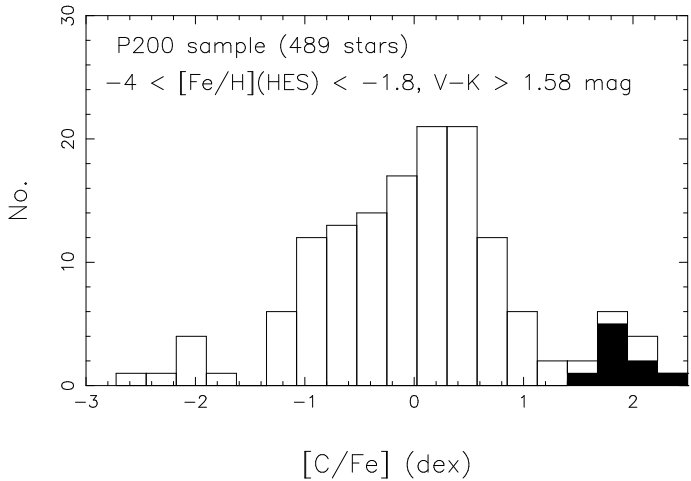}
  \caption{A histogram of [C/Fe] is shown for the giants from the P200 sample
of candidate EMP stars with $-4 \le$ [Fe/H](HES) $\le -1.8$ dex.
The known C-stars are denoted by solid fill.  The predictions of
M. Briley described in \cite{m15} for the strength of the G band of
CH based on synthetic spectra were used.}\label{fig_cfe}
\end{figure}

We next establish the distribution of C abundances and C/Fe ratios for
those stars that are not classified as C-stars.  If [C/Fe] $\ge +1.0$ dex,
we call such stars C-enhanced. Fig.~\ref{fig_ch_vmk}
displays the GP (G band of CH) index as a function of V$-$K color for the
Palomar sample.  The entire set of known C-stars from our database is shown;
for the giants they are concentrated at the top of the distribution, having
the strongest GP indices at each color.
To determine the C abundances we use the
predicted CH band strengths computed by M. Briley
described in \cite{m15}.  These were calculated
using a full spectral synthesis for the GP index  defined
in \cite{beers99}.  $V-K$ is used to define \teff, adopting a mean
value of E(B$-$V) of 0.05 mag for the HES stars. Fig.~\ref{fig_cfe} shows the resulting histogram of
[C/Fe] for the giants from this sample with [Fe/H] $\le -1.8$ dex.

The histogram suggests a peak near [C/Fe] $\sim$ 0.3 dex,
somewhat higher than the normal
value for unmixed very metal poor giants of [C/Fe] = 0.0 dex given by \cite{m15}.
There is a sharp drop towards higher C/Fe ratios, with the C-stars occupying the
upper limit of the range near [C/Fe] $\sim$ +2.0 dex.  The distribution
slowly decreases for C/Fe ratios lower than the peak.

The fraction of  giants that are C-enhanced but are not known C-stars is
6.5$\pm2.7$\%.

\section{Abundance Analysis of Carbon Stars From the HES}

To establish the results presented above, we
have carried out a detailed abundance analysis using
high dispersion spectra from HIRES at Keck
for a sample of 14 of the 16 carbon stars found in our database
among candidate EMPstars selected from the HES.
Using a \teff\ scale based on
V-I, V-J and V-K colors, we find that the Fe-metallicities
for the cooler C-stars (\teff\ $\sim$5100~K)
have been underestimated by a factor of $\sim$10 by the standard
HES survey tools.
Since we have not used the high resolution spectra themselves to determine
\teff\ or \grav, the ionization equilibrium is a stringent test of our analysis
and procedures, including the assumption of LTE.
We obtain a mean for
log$\epsilon$(Fe:Fe~II) $-$ log$\epsilon$(Fe:Fe~I) of $-0.08$ dex,
with a 1$\sigma$ rms scatter about the mean of 0.18 dex.
C-enhancement in these very metal poor C-stars appears
to reach a maximum just below the solar $\epsilon$(C), and
shows evidence of decreasing with decreasing \teff\ (increasing
luminosity), presumably due to mixing and dredge-up of C-depleted material.

\begin{figure}
\includegraphics[width=3.4in]{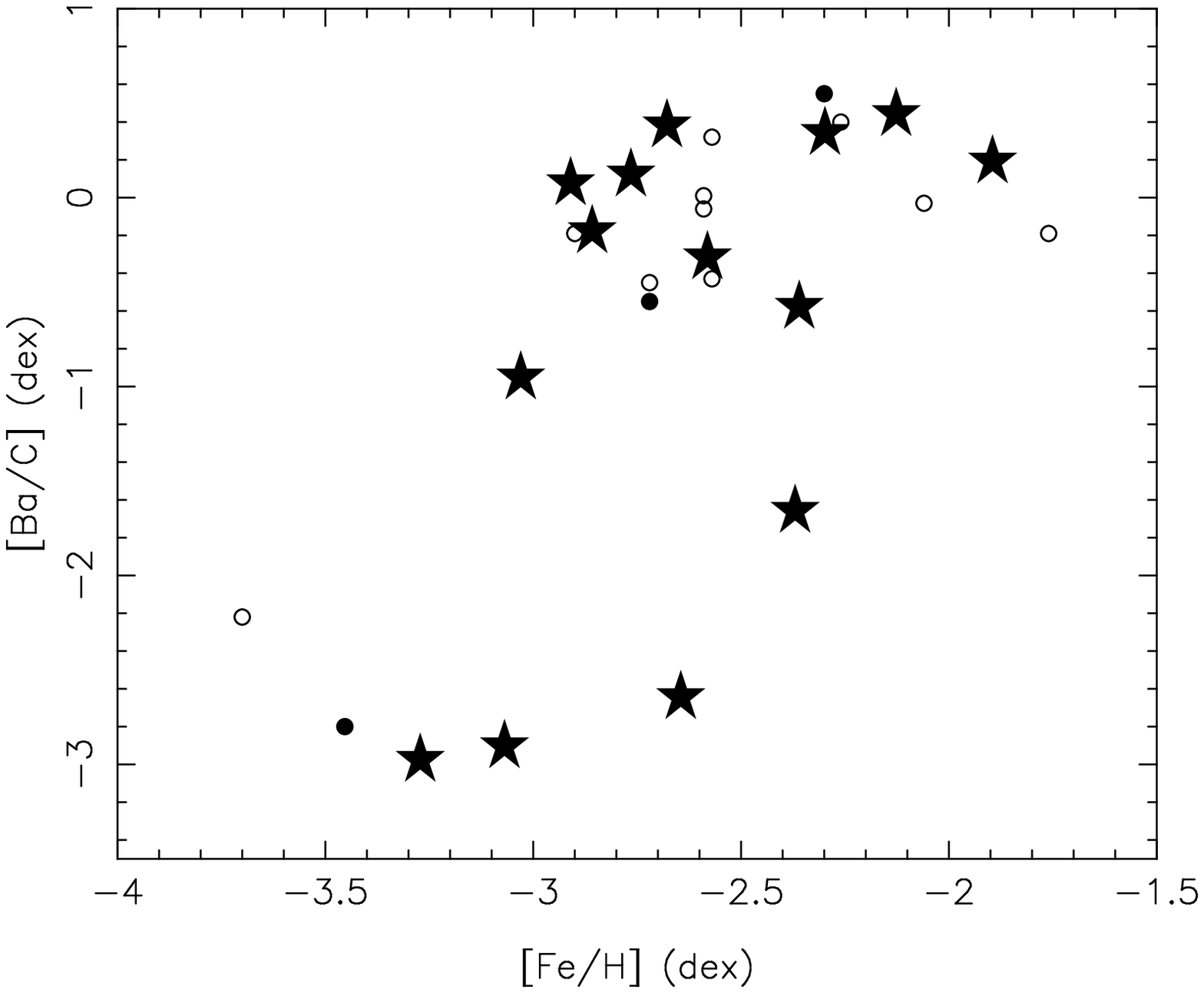}
  \caption{[Ba/C] is shown as a function of [Fe/H] for a sample
  of 24 C-stars (14 from the present sample, denoted by
  filled stars, 10 from the literature, indicated by open
  circles) and 2 C-enhanced dwarfs (filled circles).}\label{fig_ba_feh}
\end{figure}

Fig.~\ref{fig_ba_feh} shows [Ba/C] as a function of Fe-metallicity for this
sample of C and C-enhanced stars.  
Ten of the C-stars from the HES that we have analyzed
show an enhancement of Ba (and of the other $s$-process
neutron capture heavy elements) approximately equal to that of C.
The other four show [Ba/C] $\leq -1.2$ dex, i.e. a strong
C enhancement, with more normal heavy elements, as contrasted to
enhancement of both C and the $s$-process elements in the majority
of the C-stars.  Including 10
additional C-stars compiled from the literature,
$\sim$80\% of
the full sample of EMP/VMP C-stars show highly enhanced Ba,
while $\sim$20\%  has [Ba/C] $\le -1.2$ dex.
It is clear from both
the very high enhancements of lead and the Ba/Eu ratios seen among
these C-stars that the $s$-process is responsible for the enhancement
of the heavy neutron-capture elements, when present.

At first sight, this behavior suggests that two distinct processes are
required to produce the Ba-enhanced C-stars and those that are not.
Nucleosynthesis
within an intermediate mass AGB star can reproduce the first set of characteristics,
but these C-stars  are in general of too low a mass and are not sufficiently
luminous or evolved to be AGB stars.  We suggest instead that
the observed star is
the former secondary in a binary system across which
mass transfer has occured.

What about the $\sim$20\% of the C-stars without heavy element enhancements ?
We suggest that there is no need to resort to intrinsic production or any
other additional mechanism, and that essentially {\it{all}} of these stars are the
original secondary stars in mass transfer binary systems.  We ascribe
the differing enhancement of the $s$-process elements
from C-star to C-star within our sample
to some dependence in the nucleosynthetic yields
involving, for example, the initial [Fe/H] and
mass of the original primary star.
At the lowest metallicities,
\cite{busso99} predict that little or no $s$-process enhancement
will occur at the Ba peak as there will be so few Fe-seed nuclei that they
will essentially all end up as lead.  Fig.~\ref{fig_ba_feh}
provides some support for this prediction.

In our hypothesis, essentially all of these C-stars were once binaries. Assuming
most have survived to date without being disrupted, they should still be
binaries with (invisible) white dwarf companions.  The statistics
of binary detection among very metal poor C-stars from the HES
cannot be evaluated from our sample
as most of the stars have only been observed at a single epoch.
So we look instead at the sample of C-stars added from the literature,
which are mostly from the HK Survey and are
in general brighter than the HES C-stars in our sample.
They have been known as interesting objects for timescales
of several years to a decade, giving more opportunity for radial
velocity monitoring. Four of these 11 C-stars are confirmed binaries with
measured periods,
consistent with the preliminary results of
the $v_r$ monitoring program of
\cite{tsangarides} for $s$-process enhanced C-stars.
Although the sample is small,
considering the lack of suitable long-term radial velocity  monitoring programs,
the length of the typical period, the small velocity amplitudes,
the faintness of the stars, and the relatively short time they have
been known to be interesting, we find our detection rate for binaries
among very metal poor and EMP C-strs to be consistent with
all such stars being binaries;
simulations by \cite{lucatello} support this. 
There is as yet insufficient $v_r$ monitoring data for the much smaller sample
of C-enhanced but not $s$-process enhanced stars.
\cite{cohen_cabund} will present full details.

\begin{acknowledgments}
We are very greatful to M. Briley for
slightly extending his published grid of predicted CH indices for
EMP/VMP stars.
JGC and JM are grateful for partial support from  NSF grant
AST-0205951.  JGC is grateful for support from
the Ernest Fullam Award of the Dudley Observatory which helped
initiate this work.
The work of NC and FJZ is supported by Deutsche
   Forschungsgemeinschaft (grants Ch~214/3 and Re~353/44). NC
   acknowledges support through a Henri Chretien International
   Research Grant administered by the American Astronomical
   Society.
\end{acknowledgments}

\end{document}